\documentclass[twocolumn,showpacs,preprintnumbers,amsmath,amssymb]{revtex4}
\usepackage{tabularx,graphicx}\begin{document}
\newcommand{\beq}{\begin{equation}}
\newcommand{\eeq}{\end{equation}}
\newcommand{\beqn}{\begin{eqnarray}}
\newcommand{\eeqn}{\end{eqnarray}}
\newcommand{\bmath}{\begin{subequations}}
\newcommand{\emath}{\end{subequations}}
\title{An index to quantify an individual's scientific research output}
\author{J. E. Hirsch }
\address{Department of Physics, University of California, San Diego\\
La Jolla, CA 92093-0319}

\begin{abstract} 
I propose the index $h$, defined as the number of papers with citation number higher or equal to $h$, as a useful index to characterize the scientific output of a researcher.
\end{abstract}
\maketitle 

For the few scientists that earn a Nobel prize, the impact and relevance of their research work is unquestionable. 
Among the rest of us, how does one quantify the cumulative impact and relevance of an individual's scientific research output? In a world of not unlimited
resources such quantification (even if potentially distasteful) is often needed
for evaluation and comparison purposes, eg for university faculty recruitment and
advancement, award of grants, etc.

The publication record of an individual and the citation record are clearly data that
contain useful information. That information includes the number ($N_p$) of papers
published over $n$ years, the number of citations ($N_c^j$) for each paper ($j$), the
journals where the papers were published and their impact parameter, etc. This is a
large amount of information that will be evaluated with different criteria by
different people. Here I would like to propose a single number, the "$h$-index",
as a particularly simple and useful way to characterize the scientific
output of a researcher.

{\it A scientist has index h if h of his/her $N_p$ papers have at least h citations
each, and the other $(N_p-h)$ papers have no more than h citations each.}

The research reported here concentrated on physicists, however I suggest that the $h-$index  should be
useful for other scientific disciplines as well. (At the end of the paper I discuss some observations for the $h-$index in biological sciences.)
 The highest $h$ among physicists appears to be E. Witten's,
  $h=110$.
That is, Witten has written $110$ papers with at least $110$ citations each. That gives a lower bound on the total number of citations to Witten's papers
at $h^2=12,100$. Of course the total number of citations ($N_{c,tot}$) will 
usually be much larger than $h^2$, since $h^2$ both underestimates the
total number of citations of the $h$ most cited papers and ignores the papers
with fewer than $h$ citations. The relation between $N_{c,tot}$ and $h$ will 
depend on the detailed form of the particular distribution\cite{paper1,paper2}, and it is
useful to define the proportionality constant $a$ as
\beq
N_{c,tot}=a h^2 .
\eeq
I find empirically that $a$ ranges between $3$ and $5$.

Other prominent physicists with high $h$'s are A.J. Heeger ($h=107$), M.L. Cohen ($h=94$), A.C. Gossard ($h=94$),
P.W. Anderson ($h=91$), 
S. Weinberg ($h=88$), M.E. Fisher ($h=88$), M. Cardona ($h=86$), P.G. deGennes ($h=79$), J.N. Bahcall ($h=77$), 
Z. Fisk ($h=75$), D.J. Scalapino ($h=75$), G. Parisi ($h=73$),
S.G. Louie ($h=70$), R. Jackiw ($h=69$), F. Wilczek ($h=68$), 
C. Vafa ($h=66$), M.B. Maple ($h=66$), D.J. Gross ($h=66$), 
 M.S. Dresselhaus ($h=62$), S.W. Hawking ($h=62$).
 
I argue that $h$ is preferable to other single-number criteria commonly used to
evaluate scientific output of a researcher, as follows:

 (0) Total number of papers ($N_p$): Advantage: measures productivity. Disadvantage: does not measure importance nor impact of papers.

(1) Total number of citations ($N_{c,tot}$): Advantage: measures total impact. 
Disadvantage: hard to find; may be inflated by a small number of 'big hits',
which may not be representative of the individual if he/she is coauthor with
many others on those papers. In such cases the relation Eq. (1) will imply a 
very atypical value of $a$, larger than $5$.  Another disadvantage
is that $N_{c,tot}$ gives undue weight to highly cited review articles versus original
research contributions.

(2) Citations per paper, i.e. ratio of $N_{c,tot}$ to $N_p$: Advantage: allows
comparison of scientists of different ages. Disadvantage: hard to find; rewards
low productivity, penalizes high productivity.

(3) Number of 'significant papers', defined as the number of papers with more than 
$y$ citations, for example $y=50$. Advantage: eliminates the disadvantages
of criteria (0), (1), (2), gives an idea of broad and sustained impact. Disadvantage:
$y$ is arbitrary and will randomly favor or disfavor individuals; $y$ needs to be
adjusted for different levels of seniority.

(4) Number of citations to each of the $q$ most cited papers, for example 
$q=5$. Advantage: overcomes many of the disadvantages of the criteria
above. Disadvantage: it is not a single number, making it more difficult to
obtain and compare. Also, $q$ is arbitrary and will randomly favor and
disfavor individuals.

Instead, the proposed $h$-index measures the broad impact of an individual's
work; it avoids all the disadvantages of the criteria listed above; it usually can
be found very easily, by ordering papers by 'times cited' in the Thomson ISI Web of Science database\cite{clarify};
it gives a ballpark estimate of the total number of citations, Eq. (1).

Thus I argue that two individuals with similar $h$ are comparable in terms of their
overall scientific impact, even if their total number of papers or their total number
of citations is very different. Conversely, that between two individuals 
(of the same scientific age) with
similar number of total papers or of total citation count and very different
$h$-values, the one with the higher $h$ is likely to be the more accomplished
scientist.

For a given individual one expects that $h$ should increase approximately linearly with time.   
 In the simplest possible model, assume the researcher publishes $p$ papers per
year and each published paper earns $c$ new citations per year every 
subsequent year. The total number of citations after $n+1$ years is then
\beq
N_{c,tot}=\sum_{j=1}^{n}pcj=\frac{pcn(n+1)}{2}
\eeq
Assuming all papers up to year $y$ contribute to the index $h$ we have
\bmath
\beq
(n-y)c=h
\eeq
\beq
py=h
\eeq
\emath
The left side of Eq. (3a) is the number of citations to the most recent of the papers contributing to $h$; the left side of Eq. (3b) is the total
number of  papers contributing to $h$. Hence from Eq. (3),
\beq
h=\frac{c}{1+c/p}n
\eeq
The total number of citations (for not too small $n$) is then approximately
\beq
N_{c,tot}\sim \frac{(1+c/p)^2}{2c/p}h^2
\eeq
of the form Eq. (1). The coefficient $a$ depends on the number of papers and 
the number of citations per paper earned per year as given by eq. (5). As stated earlier we find empirically that $a\sim 3$  to $5$ are
typical values. The linear relation
\beq
h\sim mn
\eeq
should  hold quite generally for scientists that produce papers of similar quality
at a steady rate over the course of their
careers, of course $m$ will  vary widely among different researchers.
In the simple linear model, $m$ is related to $c$ and $p$ as given by eq. (4). Quite generally, 
the slope of $h$ versus $n$, the parameter $m$, should provide a useful yardstick to compare scientists
of different seniority.

In the  linear model, the minimum value of $a$ in Eq. (1) is $a=2$, for the case $c=p$, 
where the papers with more than $h$ citations and those with less than $h$ citations contribute equally to
the total $N_{c,tot}$. The value of $a$ will be larger for both   $c>p$ and $c<p$. For $c>p$, most contributions to the total number of citations 
arise from the 'highly cited papers' (the $h$ papers that have $N_c>h$), while for $c<p$ it is the
sparsely cited papers (the $N_p-h$ papers that have fewer than $h$ citations each) that give the largest
contribution to  $N_{c,tot}$. We find that the first situation holds in the vast majority, if not all, cases.
For the linear model defined in this
example, $a=4$  corresponds to $c/p=5.83$ (the other value that yields $a=4$, $c/p=0.17$, is
 unrealistic).

\begin{figure}
\resizebox{6.5cm}{!}{\includegraphics[width=7cm]{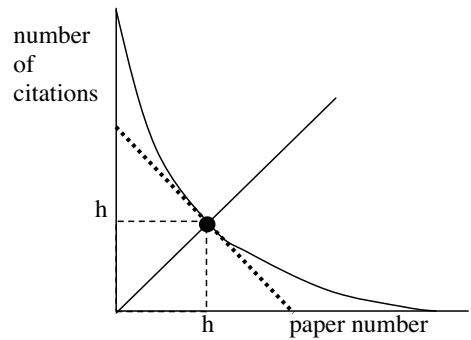}}
\caption{
The intersection of the 45 degree line with the curve giving the number of citations versus the paper number gives h.
The total number of citations is the area under the curve. Assuming the second derivative is non-negative 
everywhere, the minimum area is given by the distribution indicated by the dotted line, yielding a=2 in Eq. 1.}
\label{Fig. 1}
\end{figure}

The linear model defined above corresponds to the distribution
\beq
N_c(y)=N_0-(\frac{N_0}{h}-1)y
\eeq
where $N_c(y)$ is the number of citations to the $y$-th paper (ordered from most to least cited), and $N_0$ is the
number of citations of the most highly cited paper ($N_0=cn$ in the example above). The total number of papers
$y_m$ is given by $N_c(y_m)=0$, hence
\beq
y_m=\frac{N_0h}{N_0-h}
\eeq
We can write $N_0$ and $y_m$ in terms of $a$ defined in Eq. (1) as
:
\bmath
\beq
N_0=h[a\pm \sqrt{a^2-2a}]
\eeq
\beq
y_m=h[a\mp \sqrt{a^2-2a}]
\eeq
\emath
For $a=2$, $N_0=y_m=2h$. For larger $a$, the upper sign in Eq. (9) corresponds to the case where the highly cited papers
dominate (more realistic case) and the lower sign where the low-cited papers dominate the total citation count.

In a more realistic model, $N_c(y)$ will not
be a linear function of $y$. Note that $a=2$ can safely be assumed to be a lower bound quite generally, since a smaller
value of $a$ would require the second derivative $\partial^2N_c/\partial y^2$ to be negative over large regions
of $y$ which is not realistic. The total number of citations is given by the area under the $N_c(y)$ curve,
that passes through the point $N_c(h)=h$. In the linear model the lowest $a=2$ corresponds to 
the line of slope $-1$, as shown in Figure 1. 

A more realistic model would be a stretched exponential of the form
\beq
N_c(y)=N_0 e^{-(\frac{y}{y_0})^\beta} .
\eeq
Note that for $\beta\leq 1$, $N_c''(y)>0$ for all $y$, hence $a>2$ is true. We can write the distribution in terms
of $h$ and $a$ as
\beq
N_c(y)=\frac{a}{\alpha I(\beta)}h e^{-(\frac{y}{h\alpha})^\beta}
\eeq
with $I(\beta)$ the integral
\beq
I(\beta)=\int_0^\infty dz e^{-z^\beta}
\eeq
and $\alpha$ determined by the equation
\beq
\alpha e^{\alpha^{-\beta}}=\frac{a}{I(\beta)}
\eeq
The maximally cited paper has citations
\beq
N_0=\frac{a}{\alpha I(\beta)} h
\eeq
and the total number of papers (with at least one citation) is determined by $N(y_m)=1$ as
\beq
y_m=h[1+\alpha^\beta ln( h)]^{1/\beta}
\eeq

A given researcher's distribution can be modeled by choosing the most appropriate $\beta$ and $a$ for that case.
For example, for $\beta=1$, if $a=3$, $\alpha=0.661$ and $N_0=4.54 h$, $y_m=h[1+.66 lnh]$. With $a=4$,
$\alpha=0.4644$, $N_0=8.61h$ and $y_m=h[1+0.46 ln(h)]$. For $\beta=0.5$, the lowest possible value of
$a$ is $3.70$; for that case, $N_0=7.4 h$, $y_m=h[1+0.5 ln(h)]^2$. Larger $a$ values will increase $N_0$ and
reduce $y_m$. For $\beta=2/3$, the smallest possible $a$ is $a=3.24$, for which case
$N_0=4.5 h$ and $y_m=h[1+0.66 ln (h)]^{3/2}$.

The linear relation between $h$ and $n$ Eq. (6) will of course break down when the researcher slows down in paper
production or stops publishing altogether. There is a time lag between the two events. In the linear model
 assuming the researcher stops publishing after $n_{stop}$ years, $h$ continues to increase at the
same rate for a time
\beq
n_{lag}=\frac{h}{c}=\frac{1}{1+c/p} n_{stop}
\eeq
and then stays constant, since now all published papers contribute to $h$. In a more realistic model $h$ will
smoothly level off as $n$ increases rather than with a discontinuous change in slope. Still quite generally the
time lag will be larger for scientists who have published for many years as Eq. (16) indicates.

Furthermore in reality of course not all papers will eventually contribute to $h$. Some papers with low citations will
never contribute to a researcher's $h$, especially if written late in the career when $h$ is already appreciable.
As discussed by Redner\cite{paper3}, most papers earn their citations over a limited period of popularity and
then they are no longer cited. Hence it will be the case that papers that contributed to a researcher's $h$ 
early in his/her career will no longer contribute to $h$ later in the individual's career. Nevertheless it is of course
always true that $h$ cannot decrease with time. The paper or papers that at any given time have exactly
$h$ citations are at risk of being eliminated from the individual's $h$-count as they are superseded by other 
papers that are being cited at a higher rate.
 It is also possible that papers 'drop out' and then later come back into the $h$-count, as would occur
for the kind of papers termed 'sleeping beauties'\cite{paper4}.

For the individual researchers mentioned earlier I find n from the time elapsed since
their first published paper till the present, and find the following values for the slope $m$ defined in
Eq. (6): Witten, $m=3.89$; Heeger, $m=2.38$; Cohen,
$m=2.24$; Gossard, $m=2.09$; Anderson, $m=1.88$; Weinberg, $m=1.76$; Fisher, $m=1.91$; Cardona, $m=1.87$; deGennes,  $m=1.75$; 
Bahcall, $m=1.75$; Fisk, $m=2.14$; Scalapino, $m=1.88$;  Parisi, $m=2.15$; 
Louie, $m=2.33$; Jackiw, $m=1.92$; Wilczek, $m=2.19$; 
Vafa, $m=3.30$; Maple, $m=1.94$; Gross, $m=1.69$;  Dresselhaus, $m=1.41$; Hawking,
$m=1.59$. From inspection of the
citation records of many physicists I conclude:
 
(1) A value $m\sim1$, i.e. an $h$ index of $20$ after $20$ years of scientific
activity, characterizes a successful scientist.

(2) A value $m\sim 2$, i.e. an $h$-index of $40$ after $20$ years of scientific
activity, characterizes outstanding scientists, likely to be found only at the top
 universities or major research laboratories.

(3) A value $m\sim 3$ or higher, i.e. an $h$-index of $60$ after $20$ years,
or $90$ after $30$ years, characterizes truly unique individuals.

The $m$-parameter  ceases to be useful if a scientist does not
maintain his/her level of productivity, while the $h$-parameter remains
useful as a measure of cumulative achievement that may continue to increase
over time even long after   the scientist has stopped publishing altogether.

Based on typical $h$ and $m$ values found, I suggest that (with large error bars) for faculty at major
research universities $h\sim 10$ to $12$ might be a typical value for
advancement to tenure (associate professor), and $h\sim 18$ for 
advancement to full professor. Fellowship in the American Physical Society
might occur typically for $h\sim 15$ to $20$.  Membership in the US National Academy of Sciences
may  typically be associated with $h\sim 45$ and higher except in
exceptional circumstances. Note that these estimates correspond roughly to
typical number of years of sustained research production assuming an $m\sim 1$ value,
the time scales of course will be shorter for scientists with higher $m$ values. Note that the time estimates are taken from the publication of the first paper which 
typically occurs some years before the Ph.D. is earned.

There are however a number of caveats that should be kept in mind. Obviously a single number can never give more than a
rough approximation to an individual's multifaceted profile, and many other factors should be
considered in combination in evaluating an individual.  This and the fact that there can always be 
exceptions to rules should be kept in mind
especially in life-changing decisions such as the granting or denying of tenure. There will be differences in typical $h$-values in different fields, determined in part by the average number of references in a paper in the field, the average number of papers produced by each scientist in the field,
 and also by the size (number of scientists) in the field (although to
a first approximation in a larger field there are more scientists to share a larger number of citations, so typical $h$-values should not necessarily be
larger). 
Scientists working in non-mainstream areas will not achieve the same very high $h$ values as the top echelon   of 
those working  in highly topical areas. 
While I argue that a high $h$ is a reliable indicator of high
accomplishment, the converse is not necessarily always true. There is considerable variation in the 
skewness of citation distributions even within a given subfield, and   for an author with relatively low $h$ that has a  few seminal papers with 
extraordinarily high
citation counts, the $h$-index will not fully reflect that scientist's accomplishments.
 Conversely, a scientist with a high $h$ achieved mostly through papers with many coauthors would be treated overly
kindly by his/her $h$. 
Subfields with typically large collaborations (eg high energy experiment) will typically exhibit larger $h$-values, and
I suggest that in cases of large differences in the number of coauthors it may be useful in comparing different individuals to normalize $h$ by a factor that reflects the average number of coauthors. For determining the scientific 'age' in the computation of $m$, the very first paper may sometimes not be the 
appropriate starting point if it represents a relatively minor early contribution well before sustained productivity ensued.

Finally, in any measure of citations ideally one would like to eliminate the self-citations. While self-citations can obviously increase a scientist's $h$,
 their  effect on $h$  is much smaller than on  the total citation count. First, all self-citations to papers with less than $h$ citations are irrelevant,
 as are the self-citations to papers with many more than $h$ citations. To correct $h$ for self-citations one would consider the papers with number of citations just above $h$, and count the number of self-citations in each. If a paper with $h+n$ citations has more than $n$ self-citations, it would be dropped
 from the $h$-count, and $h$ would drop by $1$. Usually this procedure would   involve only very few if any papers. As the other face of this coin, scientists intent in increasing their $h$-index by self-citations would naturally target those papers with citations just below $h$.

\begin{figure}
\resizebox{6.5cm}{!}{\includegraphics[width=7cm]{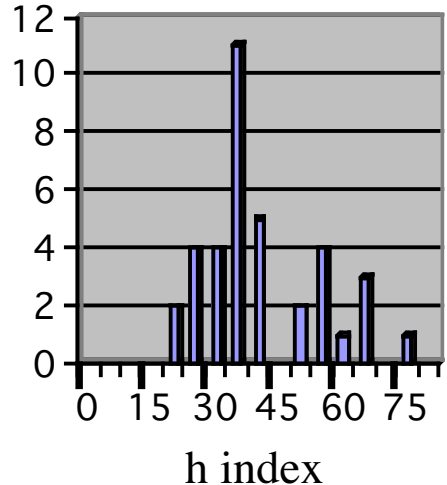}}
\caption{
Histogram giving number of Nobel-prize recipients in Physics in the last 20 years versus their h-index. 
The peak is at h-index between 35 and 39.}
\label{Fig. 2}
\end{figure}

As an interesting sample population I computed $h$ and $m$ for the
physicists that obtained Nobel prizes in the last 20 years (for calculating $m$ I used the latter of 
the first published paper year or 1955, the first year in the ISI database). However the set
was further restricted by including only the names that uniquely identified the
scientist in the ISI citation index. This restricted our set to $76\%$ of the  
total, it is however still an unbiased estimator since the commonality of the
name should be uncorrelated with $h$ and $m$. $h$-indices range from $22$ to $79$,
$m$-indices from $0.47$ to $2.19$. Averages and standard deviations are 
$<h>=41 $, $\sigma _h=15$, and $<m>=1.14$, $\sigma _m=0.47$. The distribution of $h$-indices is shown in
Figure 2, the median is at $h_m=35$, lower than the 
mean due to the tail for high $h$ values. It is interesting that Nobel prize winners  
have substantial $h$ indices ($84\%$ had $h$ of at least $30$), indicating that 
Nobel prizes do not originate in one stroke of luck but in a body of scientific work.
Notably the values of $m$ found are often not high compared to other successful scientists ($49\%$ of our
sample had $m<1$).
This is clearly because Nobel prizes are often awarded long after the
period of maximum productivity of the researchers.

As another example, among newly elected members in the National Academy of Sciences in Physics and Astronomy in 2005 I find 
$<h>=44$, $\sigma_h=14$, highest $h=71$, lowest $h=20$, median $h_m=46$. Among the total membership in NAS in Physics 
the subgroup of last names starting with A and B has 
$<h>=38$, $\sigma_h=10$, $h_m=37$. These examples further indicate that the index $h$ is a stable and consistent estimator of
scientific achievement.

An intriguing idea is the   extension of the $h$-index concept to groups of individuals\cite{spires0}. The SPIRES high energy physics
literature database\cite{spires} recently implemented the $h$-index in their citation summaries, and it also allows the
computation of $h$ for groups of scientists.  The overall $h$-index of a group will 
generally be larger than that of each of the members of the group but
smaller than the sum of the individual $h$-indices, since some of the papers that contribute to each individual's $h$ will no longer
contribute to the group's $h$. For example, the overall $h$-index of the condensed matter group at the UCSD physics department
is $h=118$, of which the largest individual contribution is $25$;  the highest individual $h$ is $66$, and the sum of individual $h$'s is above $300$. The contribution of each
individual to the group's $h$ is not necessarily proportional to the individual's $h$, and the highest contributor to the
group's $h$ will not necessarily be the individual with highest $h$. 
In fact, in principle (although rarely in practice) the lowest-$h$
individual in a group could be the largest contributor to the group's $h$.
For a prospective graduate student considering different
graduate programs, a ranking of groups or departments in his/her chosen area according to their overall $h$-index would likely be
of interest, and for administrators concerned with these issues the ranking of their departments or entire institution according
to the overall $h$ could also be of interest.

To conclude, I discuss some observations in the fields of biological and biomedical sciences. From the list compiled by Christopher King of Thomson ISI
of the most highly cited scientists in the period 1983-2002\cite{king}, I found the $h-$indices for the top 10 on that list, all in the life sciences, which are, in order of
decreasing $h$:
S.H. Snyder, $h=191$; D. Baltimore, $h=160$; R.C. Gallo, $h=154$; P. Chambon, $h=153$; 
B. Vogelstein, $h=151$; S. Moncada, $h=143$;  C.A. Dinarello, $h=138$; 
  T. Kishimoto, $h=134$; R. Evans, $h=127$; 
A. Ullrich, $h=120$. It can be seen that not surprisingly all  these highly cited researchers
also have high $h-$indices, and that high $h-$indices in the life sciences are much higher than in physics. Among 36 new inductees in the National Academy of Sciences in biological and biomedical sciences in 2005 I find
$<h>=57$, $\sigma_h=22$, highest $h=135$, lowest $h=18$, median $h_m=57$. These latter results confirm that $h-$indices in biological sciences tend to be higher than
in physics, however they also indicate that the difference appears to be much higher at the high end than on average. Clearly more research in understanding 
similarities and differences of $h-$index distributions in different fields of science would be of interest.

In summary, I have proposed an easily computable index, $h$, which gives an   estimate of the importance, significance and
broad impact of a scientist's cumulative research contributions. I suggest that
this index may  provide a useful yardstick to compare different individuals
competing for the same resource when an important evaluation criterion is
scientific achievement, in an unbiased way.

\acknowledgements 
I am grateful to  many colleagues in the UCSD Condensed Matter group  and especially Ivan Schuller for stimulating discussions on these topics
and  encouragement to publish these ideas; to the many readers that wrote with interesting comments since this paper was first posted
at the LANL ArXiv (http://arxiv.org/abs/physics/0508025), and to the referees who made constructive
suggestions, all of which led  to improvements  in the paper; and  to Travis Brooks and the SPIRES database administration for rapidly implementing the $h$-index in their
database.


\begin{references}
\bibitem{spires0} This was first introduced in the SPIRES database, Ref. 7.
\end{references}
\end{document}